\documentclass[twocolumn,twoside,slac_two]{revtex4}
\usepackage{graphicx}
\usepackage{fancyhdr}
\begin{document}

\title{Toward a New Paradigm for the Unification of Radio Loud AGN and its Connection to Accretion}

\author{Markos Georganopoulos}
\affiliation{Department of Physics, University of Maryland Baltimore County (UBMC), 1000 Hilltope Circle, Baltimore, MD 21250, USA \& \\ 
Astrophysics Science Division, NASA/GoddardSpaceFlight Center, Greenbelt, MD20771, USA}

\author{Eileen T. Meyer, Giovanni Fossati}
\affiliation{Department of Physics and Astronomy, Rice University, Houston, TX 77005, USA}

\author{Matthew L. Lister}
\affiliation{Department of Physics, Purdue University, West Lafayette, IN 47907, USA}

\begin{abstract}
We  recently argued \citep{mey11} that the collective properties of radio loud active galactic nuclei point to the existence of two families of sources, one of powerful sources with single velocity jets and one of weaker jets with significant velocity gradients in the radiating plasma.
These families  also correspond to different  accretion modes and therefore different thermal and emission line intrinsic properties:
powerful sources have radiatively efficient accretion disks, while in weak sources
accretion must be radiatively inefficient. 
Here, after we briefly review of our recent work, we present the following findings that support our 
unification scheme: ($i$) along the broken sequence of aligned objects, the jet kinetic power increases. ($ii$) in the powerful branch of the sequence of aligned objects the fraction of BLLs decreases with increasing jet power. ($iii$) for powerful sources, the fraction of BLLs increases for more un-aligned objects, as measured by the core to extended radio emission. Our results are also compatible with the possibility that a given accretion power produces jets of comparable kinetic power. 
\end{abstract}

\maketitle

\thispagestyle{fancy}

\section{The blazar sequence and observations that challenge it}

{\bf The sequence.} In blazars, radio loud active galactic nuclei (AGN) with their relativistic jet axis pointing  to our line of sight,  the synchrotron peak frequency ($\nu_{peak}$) covers a wide range  ($ 10^{12} \lesssim \nu_{peak}\lesssim 10^{18}$ Hz), with BL\,Lacs (BLL, lineless  blazars) spanning the entire range and FSRQs (Flat spectrum radio quasars, sources with strong broad emission lines)
having lower $\nu_{peak}$ ($ 10^{12} \lesssim \nu_{peak}\lesssim 10^{14}$ Hz). Following \cite{abd10},  we adopt the generic terms for low, intermediate, and high {\sl synchrotron-peaking} (LSP, ISP, HSP) blazars independently
of the spectroscopic type.
\cite{fos98} found  that as the source synchrotron power $L_{peak}$ increases, $\nu_{peak}$ decreases, with  predominantly  FSRQ sources at the low $\nu_{peak}$, high $L_{peak}$ end through LSP, ISP, and finally HSP BL\,Lacs at the
low $L_{peak}$ end. They also used the sparse {\sl EGRET}  data to argue that the same reduction of the peak frequency happens in the high energy - presumably inverse Compton (IC) component - component and that the Compton dominance (the ratio of IC to synchrotron power) increases   with source power. \cite{ghi98} suggested that more efficient cooling of particles in the jets of high luminosity blazars is
responsible for the lower peak frequencies. 

 {\bf From sequence to envelope.} \cite{padovani97} and \cite{perlman98} identified
relatively powerful sources with a radio to X-ray spectral index
$\alpha_{RX}$ typical of weak sources with $\nu_{peak}$ in the X-rays. Such
sources, if confirmed, challenge the sequence. Upon close study,
however, their X-ray emission was found not to be of synchrotron
origin \citep{landt08} and as of now sources with high $L_{peak}$ - high $\nu_{peak}$ have not been found
\citep{landt06,padovani07,landt08,maraschi08}.  Sources below the
blazar sequence are expected from jets less aligned to the line of
sight. Indeed, \cite{padovani03,anton05,nieppola06} found that new sources they
identified modify the blazar sequence to an {\sl envelope}.

{\bf Challenges.} \cite{caccianiga04} found several low $L_{peak}$  - low $\nu_{peak}$ sources
that, because they have a high core dominance ($R$, ratio of core and
therefore beamed to extended and therefore isotropic radio emission),
are not intrinsically bright sources at a larger jet angle.  These
sources challenge the sequence because ($i$) both intrinsically weak
and intrinsically powerful jets can have similar $\nu_{peak}$ and ($ii$)
intrinsically weak jets can produce a wide range of $\nu_{peak}$ from
($10^{12}$ - $10^{18}$ Hz). Another challenge came from
\cite{landt08} who showed that, contrary to what is anticipated by
the sequence, high and low synchrotron peak frequency (HSP and LSP) BL
Lacertae objects (BLLs, blazars with emission line EW $W<
5 $ \AA) have similar $L_{ext}$.   These findings challenge the sequence, even after being extended to include the sources in the envelope as de-beamed analogs of the blazar sequence sources.

\subsection{The case for a critical accretion rate in radio loud AGN}

\cite{nar97} argued that at a critical value of the accretion rate $\dot m_{cr}=\dot M_{acc}/\dot M_{Edd}\sim 10^{-3}\,-\, 10^{-2}$, 
the accretion switches from a standard radiatively efficient thin disk with accretion-related emission power $L_{acc}\propto \dot m L_{Edd}$ for $\dot m> \dot m_{cr}$, to a radiatively inefficient mode  
where $L_{acc}\propto \dot m^2 L_{Edd}$.
 This critical point may be connected to the transition between Fanaroff Riley \cite[FR;][]{fanaroff74} II to FR I radio galaxies (RG) : the level of the low frequency extended radio emission (coming mostly from the radio lobes and considered to be isotropic)  that separates  FR I and FR II RG, has been shown to be a function of  the host galaxy optical magnitude \citep{led96}: the division  between FR I and FR II is at higher radio luminosities for brighter galaxies. 
\cite{ghi01} argued that, because the optical magnitude of a galaxy is related to the central black home mass \citep{mcl01} and the extended radio luminosity is  related to  the jet kinetic luminosity \citep[following the scaling of][]{wil99}, 
this division can be casted as a division in terms of the fraction of the Eddington luminosity carried by the jet: jets with kinetic luminosity
$L_{kin}  \lesssim \dot m_{cr}\, L_{Edd}$  give rise to FR I RG, while jets with  $L_{kin}\gtrsim \dot m_{cr}\, L_{Edd}$  are predominantly
FR II sources.  Interestingly, and in agreement with the unification scheme, \cite{ghi08} and \cite{ghi09} find that the same dichotomy applies to separating BLLs and FSRQ, the aligned versions of FR I and FR II respectively.
Finally, it is very intriguing that \cite{mar04} argue that there is a paucity of sources around  $\dot m \sim 0.01$.  FR I, low line excitation FR II and some high line excitation FR II were found to occupy the low $\dot m$ regime, while the high   $\dot m$ regime was occupied by high line excitation FR II, broad line radio galaxies and powerful radio quasars.

\begin{figure}
\begin{center}
\includegraphics[width=2.8in]{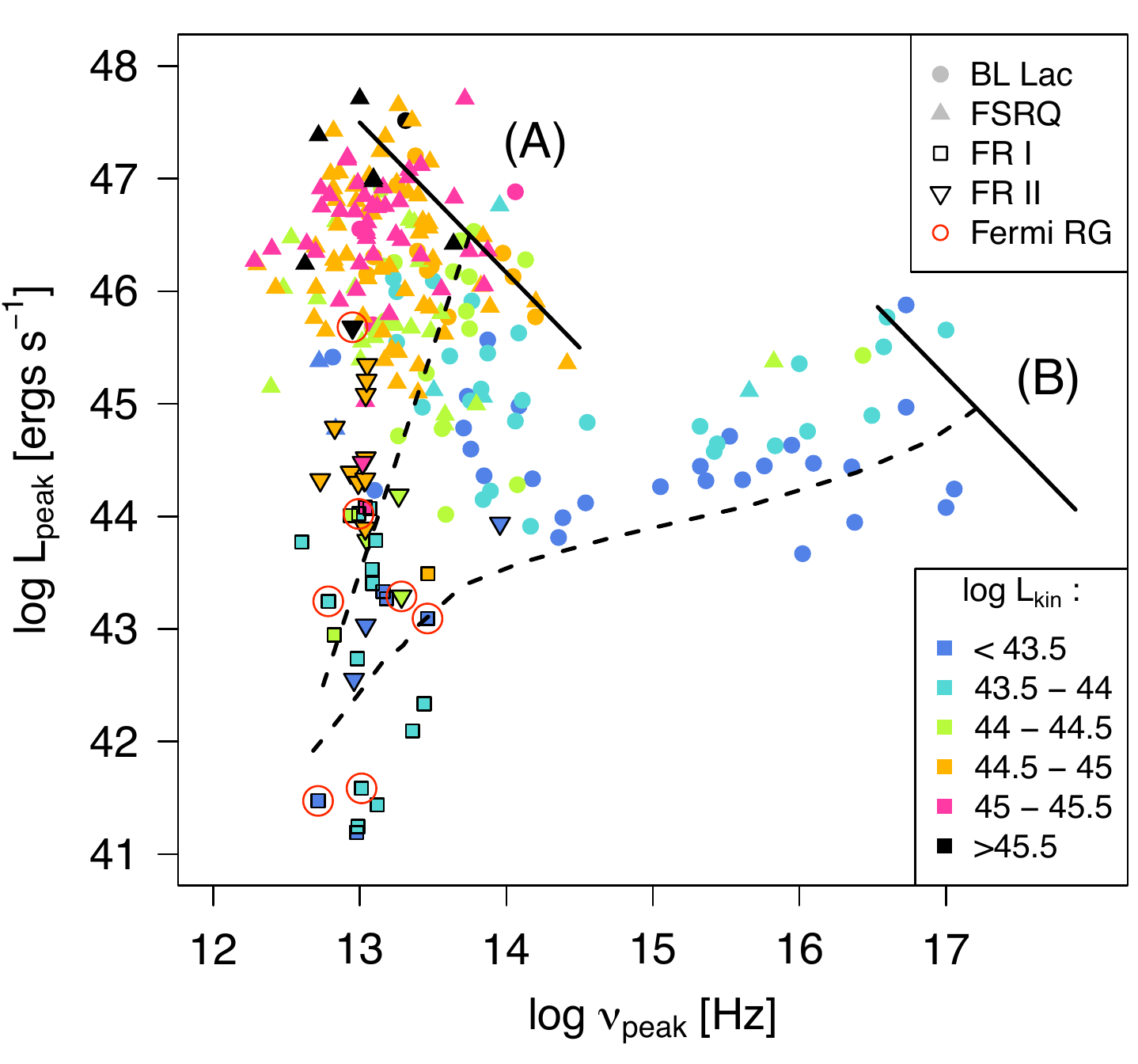}
\end{center}
\caption{\sl From M11: The blazar sequence,  has been expanded into an ``envelope'' with the addition of new  observations. The solid lines indicate the  broken power sequence of an aligned source as its $L_{kin}$ increases. 
Track (A) shows the path of a synchrotron peak for a single speed jet in an environment of radiatively efficient accretion  and (B) for a 
decelerating jet of the type hypothesized to exist in FRI sources as the jet orientation changes.}
\label{M11}
\end{figure}

\section{A broken power sequence for blazars}

Recently, we
\cite[][heretofore M11]{mey11}   compiled the largest  sample of radio loud AGN  for which sufficient data existed to determine variability-averaged  $\nu_{peak}$  $L_{peak}$, as well as the extended low frequency radio emission $L_{ext}$. This is an important quantity in our study, because it has been shown to be  a good proxy for the jet kinetic power $L_{kin}$, as measured by the energy required to inflate  the X-ray cavities seen to coincide with the radio lobes of a number of sources \citep[e.g.][]{bir08, cav10}.

The picture that emerges (figure \ref{M11}) exhibits some important differences with the blazar sequence.  In particular, ISP BLLs have  $L_{kin}$ comparable to that of HSP and LSP BLLs. Also, although all  the FR I galaxies were found to have similar $L_{kin}$ with BLLs, no FR I galaxies were found with $\nu_{peak}\gtrsim 10^{13.5} $ Hz. Because there is no obvious selection acting against the detection of FR I galaxies with core
SED peaking at higher energies, we are lead to conclude that the un-aligned versions of 
HSP blazars have $\nu_{peak}$ smaller by a factor of least $10^3$ compared to their aligned 
equivalent, something that agrees with the existence of velocity profiles in the emitting plasma, as supported by other investigations \citep[e.g.][]{chi00,tru03,geo03,ghi05}.

In M11 we suggested that {\sl  extragalactic jets can be described in terms of two families}.  The first is that of weak jets characterized by velocity profiles and weak or absent broad emission lines. HSPs ($\nu_{peak}\gtrsim10^{16}$ Hz), ISPs ($10^{14.5}\lesssim \nu_{peak}\lesssim10^{16}$ Hz), and FR I RG belong to this family. On the basis of having similar $L_{kin}$ with HSPs and FR I RG,  the ISP sources were argued to be somewhat
un-aligned HSPs. The second family  is that  of more powerful jets having  a single Lorentz factor emitting plasma  and, in most cases,  stronger broad emission lines. 
Interestingly, the two families divide  at 
$L_{kin} \sim 10^{44.5}$ erg s$^{-1}$, which for  $M=10^9\,M_\odot$, corresponds to 
   $L_{kin} \sim 2.3 \times 10^{-3} L_{Edd,9}$, similar to the $\dot m_{cr}$ of 
   \cite{nar97}.  Aligned sources are found along the broken power sequence depicted by the solid lines A and B with A (B) corresponding to jets in radiatively efficient (inefficient) accretion environments with $\dot m > \dot m_{cr}$ ($\dot m < \dot m_{cr}$).
      The broken lines A and B  depict the tracks followed by two sources as they
      depart from the power sequence and 
      their orientation angle $\theta$ increases. While in the first case a single velocity flow is assumed, in the second case emission from a decelerating flow is considered \citep{geo03}.

 \begin{figure}[t]
\centerline{
\includegraphics [width=2.8in]{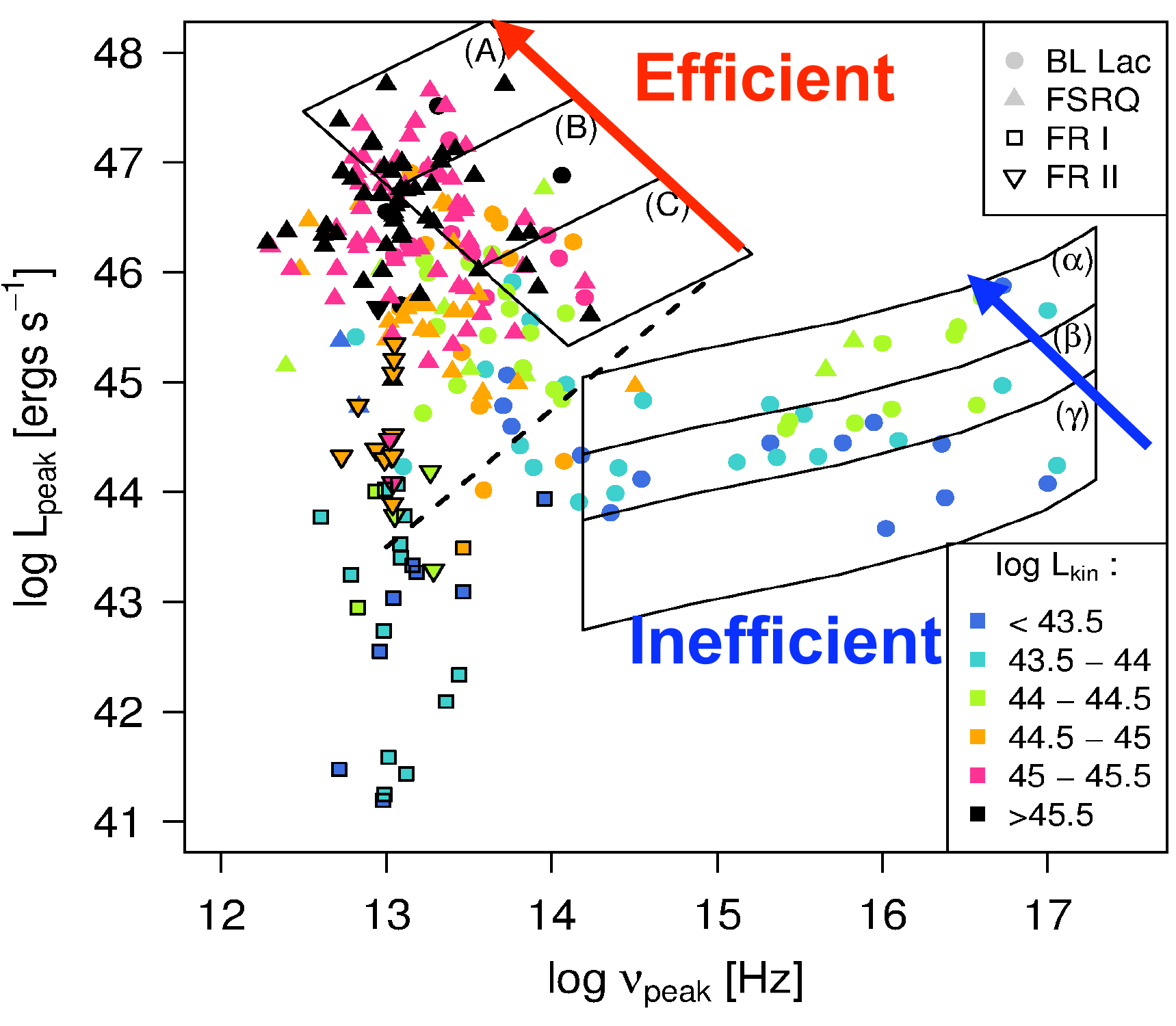}}
\caption{ \sl  Same data as Figure  \ref{M11}. See text for the  description of  boxes A, B, C and  zones $\alpha,\, \beta,\, \gamma$.}
\label{prelim}
\end{figure}

\section{Tests of the new scheme.}

We now discuss some predictions of the new unification scheme and their 
confirmation from the current data. 

\begin{figure}[t]
\centerline{
\includegraphics [width=2.7in]{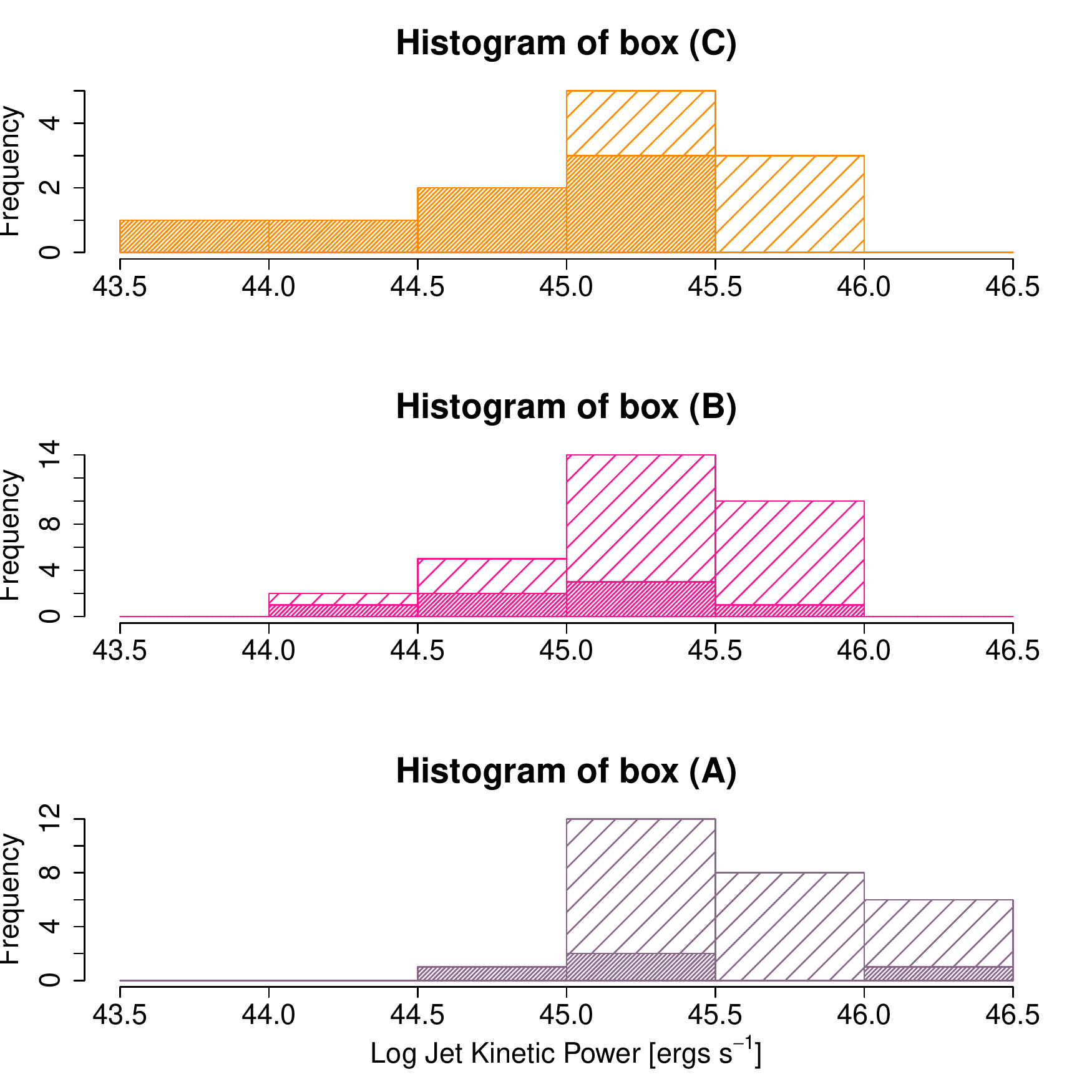}}
\caption{ \sl    Histograms of $L_{kin}$ for the sources in each box A, B, C of Figure \ref{prelim} with darker shade used for BLLs.}
\label{powerful}
\end{figure}

\begin{figure}[t]
\centerline{
\includegraphics [width=2.6in]{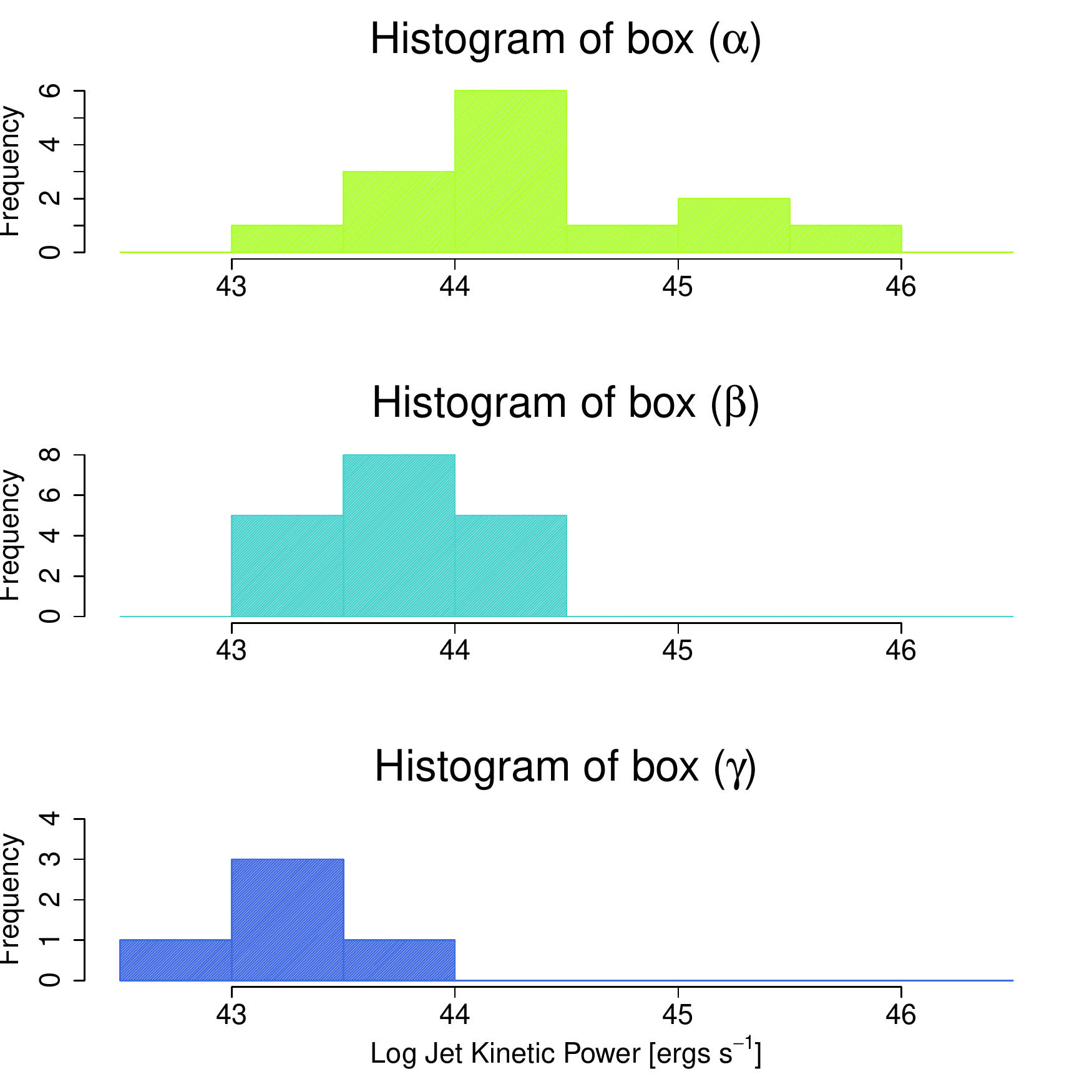}}
\caption{ \sl  Histograms of $L_{kin}$ for the sources of Figure \ref{prelim}  in each of the  zones $\alpha,\, \beta,\, \gamma$.}
\label{weak}
\end{figure}

{\bf $\bf L_{kin}$   increases  along the two branches of the broken power sequence.} 
We examine now if along the two branches of the power sequence, depicted schematically by the red and  blue arrows in the  Figure \ref{prelim}, 
$L_{kin}$ increases. To do that, we select those sources that are close to the  sequence of powerful aligned objects and split them in three groups A, B, C, as seen in Figure \ref{prelim}.
In Figure \ref{powerful} we plot the $L_{kin}$ distribution of sources in these three groups. As expected, the average $L_{kin}$ increases from group C to A. Running the same test for jets with inefficient accretion requires to use sources that are not aligned, because of the small number of sources. For this reason, we select all low power sources with $\log \nu_{peak} > 14.2$ to insure that we do not have any mixing with sources of the other branch and we separate them in the  three groups $\alpha,\,\beta,\,\gamma$ (Figure \ref{prelim}) separated by the de-beaming tracks of a decelerating jet depicted also in Figure \ref{M11}. As can be seen in  Figure \ref{weak},  the average $L_{kin}$ increases from group $\gamma$ to $\alpha$, according to our expectations.

{\bf As $\bf L_{kin}$ increases,  the fraction of the BLLs decreases along the powerful sequence.}  As $L_{kin}$ increases along the powerful sequence, $L_{peak}$ increases, but $\nu_{peak}$ decreases. At the same time, if we assume that $L_{kin}$ scales with  accretion power, we expect that the BLR luminosity increases. If $\nu_{peak}$ did not change, we would expect that the ratio of the BLR to optical synchrotron emission
would not change.  But $\nu_{peak}$ does decrease as $L_{kin}$ increases, shifting the synchrotron component to lower frequencies and revealing more of the BLR.
Thus we expect that the fraction of sources that is classified as BLLs will become smaller as $L_{kin}$ increases along the powerful sequence.  This is clearly seen in  Figure \ref{powerful}, with the fraction of BLLs clearly decreases as $L_{kin}$ decreases.

\begin{figure}[t]
\centerline{
\includegraphics [width=2.6in]{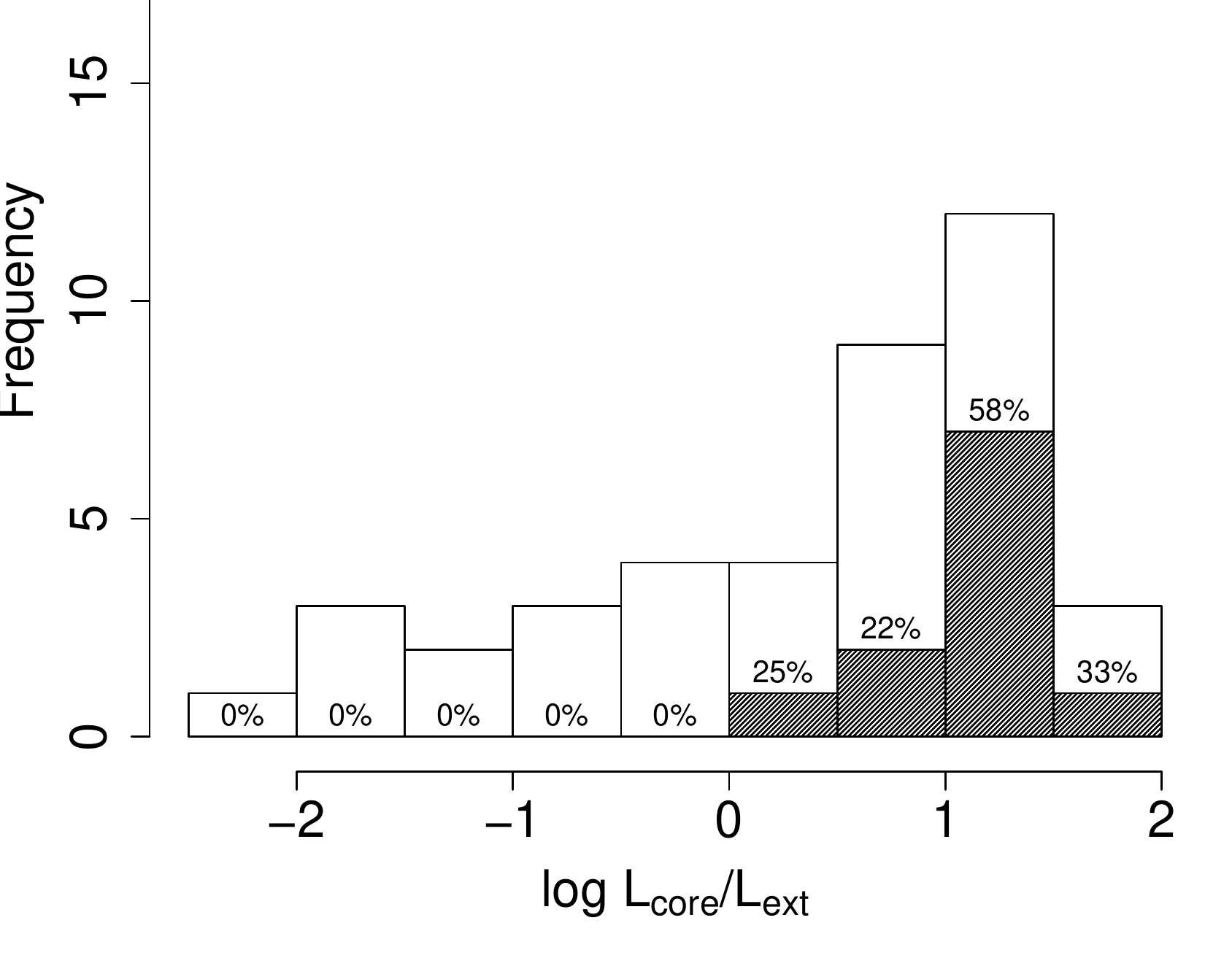}}
\caption{ \sl  Number of sources as a function of radio core dominance for all sources with $10^{44.5}<L_{kin}<10^{45}$ erg s$^{-1}$ (orange sources in Figure \ref{M11}). The dark shade corresponds to BLLs.}
\label{bll_fraction}
\end{figure}

{\bf For powerful sources, the fraction of BLLs increases for less aligned sources.}
In our scheme, we expect that for powerful sources of a given $L_{kin}$, as they become
more un-aligned, the beamed synchrotron emission will decrease, while the BLR luminosity will be much less affected, resulting to a decreasing fraction of BLLs for more un-aligned sources. To address this, we selected sources with $10^{44.5}<L_{kin}<10^{45}$ erg s$^{-1}$ (orange sources in  Figure \ref{prelim}) and we plotted the fraction of BLLs as a function of radio core dominance $L_{core}/L_{ext}$ which is an orientation indicator. As can be seen in  Figure \ref{bll_fraction}, as the core dominance decreases, the fraction of  BLLs quickly decreases, in  agreement with our expectations.

\begin{figure}[b]
\centerline{
\includegraphics [width=2.7in]{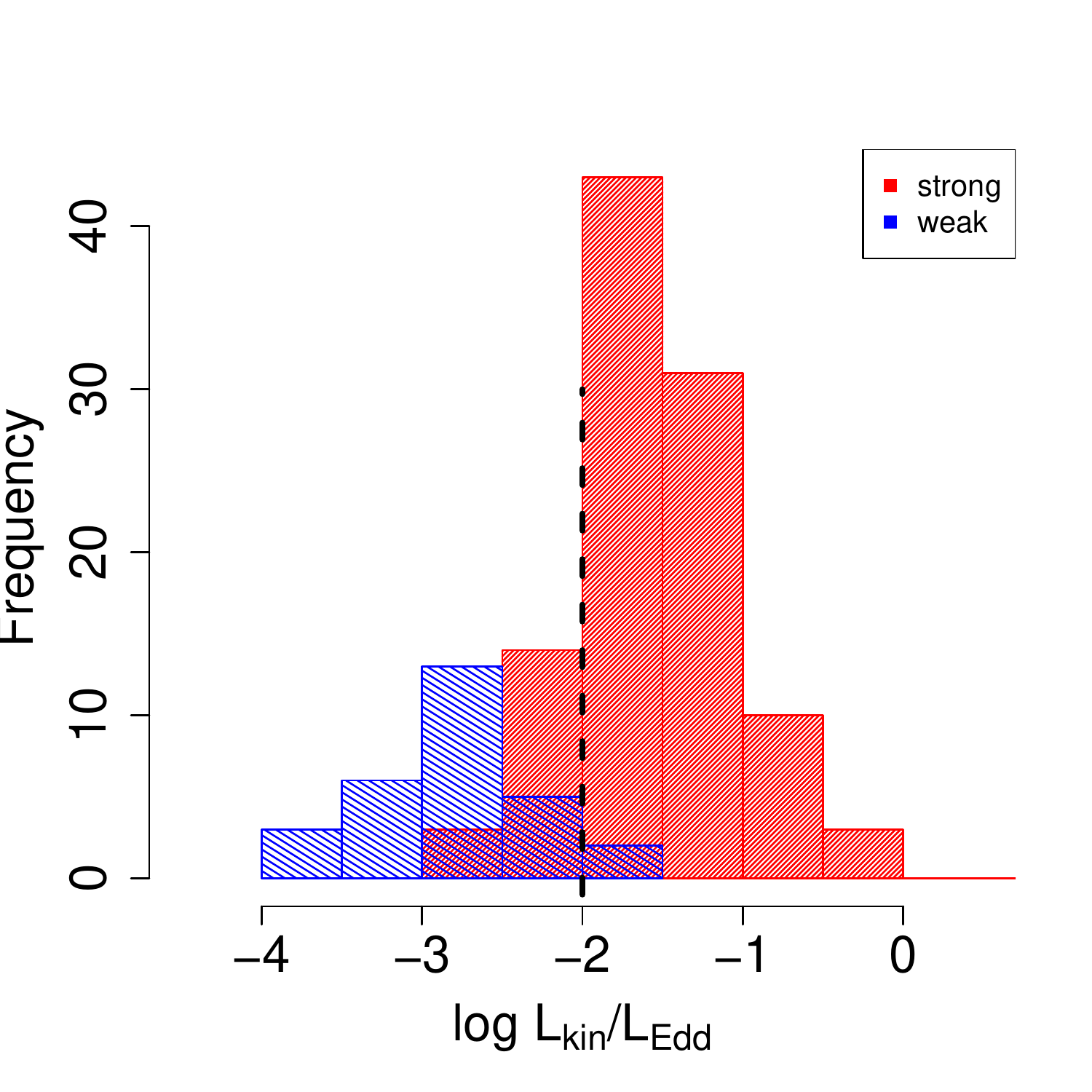}}
\caption{ \sl Distribution of $L_{kin}/L_{Edd}$ for blue (red) sources with radiatively inefficient (efficient) accretion disks.}
\label{mcrit}
\end{figure}

{\bf A given accretion power $\bf L_{acc}$  corresponds to a narrow $\bf L_{kin}$ range.} We collected black hole masses from the literature for most of the sources of M11 and used them to calculate the ratio of $L_{kin}/L_{Edd}$. We plot our results in  Figure \ref{mcrit}: in blue sources with $\nu_{peak}>10^{14.5}$ Hz, almost exclusively BLLs, therefore radiatively inefficient accretors; in red sources with 
$\nu_{s}< 10^{14.5}$ Hz and $L_{s}>10^{45.5}$ erg s$^{-1}$, almost all FSRQs, therefore radiatively efficient accretors.
The  separation of red and blue sources at $L_{kin}/L_{Edd}\sim 10^{-2}$ suggests that there is 
a transition at $\dot m=\dot m_{cr}\sim 10^{-2}$ with radiatively efficient accretion at 
$\dot m>\dot m_{cr}$ and that sources with a given accretion power do not produce jets with $L_{kin}$ significantly smaller or larger than their accretion power.





\bigskip 

\begin{thebibliography}{99}   




\bibitem[Abdo et al.(2010)]{abd10} Abdo, A.~A., Ackermann, 
M., Agudo, I., et al.\ 2010, ApJ, 716, 30 


\bibitem[Ant{\'o}n 
\& Browne(2005)]{anton05} Ant{\'o}n, S., \& Browne, I.~W.~A.\ 2005, MNRAS, 356, 225 



 \bibitem[B{\^i}rzan et al.(2008)]{bir08} B{\^i}rzan, L., 
McNamara, B.~R., Nulsen, P.~E.~J., Carilli, C.~L., 
\& Wise, M.~W.\ 2008, ApJ, 686, 859

\bibitem[Caccianiga 
\& March{\~a}(2004)]{caccianiga04} Caccianiga, A., \& March{\~a}, M.~J.~M.\ 2004, MNRAS, 348, 937 


\bibitem[Cavagnolo et al.(2010)]{cav10} Cavagnolo, K.~W.,  McNamara, B.~R., Nulsen, P.~E.~J., et al.\ 2010, ApJ, 720, 1066 



\bibitem[Chiaberge et 
al.(2000)]{chi00} Chiaberge, M., Celotti, A., Capetti, A., \& Ghisellini, G.\ 2000, A\&A, 358, 104 

\bibitem[Fanaroff 
\& Riley(1974)]{fanaroff74} Fanaroff, B.~L., \& Riley, J.~M.\ 1974, MNRAS, 167, 31P 


\bibitem[Fossati et al.(1998)]{fos98} Fossati, G., Maraschi, 
L., Celotti, A., Comastri, A., \& Ghisellini, G.\ 1998, MNRAS, 299, 433 

\bibitem[Ghisellini et al.(1998)]{ghi98} Ghisellini, G., 
Celotti, A., Fossati, G., Maraschi, L., 
\& Comastri, A.\ 1998, MNRAS, 301, 451 

\bibitem[Ghisellini 
\& Celotti(2001)]{ghi01} Ghisellini, G., \& Celotti, A.\ 2001, A\&A, 379, L1 

\bibitem[Ghisellini et 
al.(2005)]{ghi05} Ghisellini, G., Tavecchio, F., \& Chiaberge, M.\ 2005, A\&A, 432, 401 

\bibitem[Ghisellini 
\& Tavecchio(2008)]{ghi08} Ghisellini, G., \& Tavecchio, F.\ 2008, MNRAS, 387, 1669 

\bibitem[Ghisellini et al.(2009)]{ghi09} Ghisellini, G., 
Maraschi, L., \& Tavecchio, F.\ 2009, MNRAS, 396, L105 



\bibitem[Georganopoulos 
\& Kazanas(2003)]{geo03} Georganopoulos, M., \& Kazanas, D.\ 2003, ApJL, 594, L27 

\bibitem[Landt et al.(2006)]{landt06} Landt, H., Perlman, 
E.~S., \& Padovani, P.\ 2006, ApJ, 637, 183 

\bibitem[Landt et al.(2008)]{landt08} Landt, H., Padovani, P., 
Giommi, P., Perri, M., \& Cheung, C.~C.\ 2008, ApJ, 676, 87 

\bibitem[Ledlow 
\& Owen(1996)]{led96} Ledlow, M.~J., \& Owen, F.~N.\ 1996, AJ, 112, 9 



\bibitem[Maraschi et al.(2008)]{maraschi08} Maraschi, L., 
Foschini, L., Ghisellini, G., Tavecchio, F., 
\& Sambruna, R.~M.\ 2008, MNRAS, 391, 1981 

\bibitem[Marchesini et al.(2004)]{mar04} Marchesini, D., 
Celotti, A., \& Ferrarese, L.\ 2004, MNRAS, 351, 733 


\bibitem[McLure 
\& Dunlop(2001)]{mcl01} McLure, R.~J., \& Dunlop, J.~S.\ 2001, MNRAS, 327, 199 

\bibitem[Meyer et al.(2011)]{mey11} Meyer, E.~T., Fossati, 
G., Georganopoulos, M., \& Lister, M.~L.\ 2011, ApJ, 740, 98

\bibitem[Narayan et al.(1997)]{nar97} Narayan, R., Garcia, 
M.~R., \& McClintock, J.~E.\ 1997, ApJ, 478, L79 


\bibitem[Nieppola et al.(2006)]{nieppola06} Nieppola, E., Tornikoski, M., \& Valtaoja, E.\ 2006, A\&A, 445, 441 

\bibitem[Padovani et al.(1997)]{padovani97} Padovani, P., Giommi, 
P., \& Fiore, F.\ 1997, MNRAS, 284, 569 

\bibitem[Padovani et al.(2003)]{padovani03} Padovani, P., Perlman, 
E.~S., Landt, H., Giommi, P., \& Perri, M.\ 2003, ApJ, 588, 128 


\bibitem[Padovani et al.(2007)]{padovani07} Padovani, P., Giommi, 
P., Landt, H., \& Perlman, E.~S.\ 2007, ApJ, 662, 182 



\bibitem[Perlman et al.(1998)]{perlman98} Perlman, E.~S., 
Padovani, P., Giommi, P., et al.\ 1998, AJ, 115, 1253 


\bibitem[Trussoni et 
al.(2003)]{tru03} Trussoni, E., Capetti, A., Celotti, A., Chiaberge, M., \& Feretti, L.\ 2003, A\&A, 403, 889 

\bibitem[Willott et al.(1999)]{wil99} Willott, C.~J., 
Rawlings, S., Blundell, K.~M., \& Lacy, M.\ 1999, MNRAS, 309, 1017 




\end{thebibliography}

\end{document}